\documentclass[
    ,final            
  ]
  {aipproc}

\layoutstyle{6x9}
\newcommand{\lsim}   {\mathrel{\mathop{\kern 0pt \rlap
  {\raise.2ex\hbox{$<$}}}
  \lower.9ex\hbox{\kern-.190em $\sim$}}}
\newcommand{\gsim}   {\mathrel{\mathop{\kern 0pt \rlap
  {\raise.2ex\hbox{$>$}}}
  \lower.9ex\hbox{\kern-.190em $\sim$}}}
\newcommand{\K}{\textrm{K}}

\begin{document}

\title{On the dark energy rest frame and the CMB}



\author{Jos\'e Beltr\'an Jim\'enez and Antonio L. Maroto}{
  address={Departamento de F\'isica Te\'orica, Universidad Complutense de Madrid, 28040 Madrid, Spain}
}

\begin{abstract}
Dark energy is usually parametrized as a perfect fluid with
negative pressure and a certain equation of state. Besides, it is
supposed to interact very weakly with the rest of the components
of the universe and, as a consequence, there is no reason to
expect it to have the same large scale rest frame as matter and
radiation. Thus, apart from its equation of state $w$ and its
energy density $\Omega_{DE}$ one should also consider its velocity
as a free parameter to be determined by observations. This
velocity defines a cosmological preferred frame, so the universe
becomes anisotropic and, therefore, the CMB temperature
fluctuations will be affected, modifying mainly the dipole and the
quadrupole.
\end{abstract}

\maketitle


\section{Introduction}

Recent observations \cite{Riess,Astier,Spergel} suggest that the
universe could be dominated by a fluid with negative pressure
\cite{review,review2} which has been called dark energy (DE). This
DE component is usually assumed to behave as a perfect fluid with
(possibly) a time-evolving equation of state
$p_{DE}=w_{DE}(z)\rho_{DE}$ and whose interactions with the rest
of components of the universe are very weak \cite{review2}.
Therefore, apart from the density parameter $\Omega_{DE}$ and
equation of state $w_{DE}$, a complete knowledge of its
energy-momentum tensor requires the determination of its relative
velocity with respect to the rest of components of the universe
or, in other words, its large scale rest frame. In fact, recent
measurements of peculiar velocities of matter bulks with respect
to the CMB \cite{darkflow} have shown the existence of a coherent
matter flow on scales $\lsim 300h^{-1}$Mpc. These observations
suggest that matter and radiation rest frames could differ from
each other at large scales even though they were strongly coupled
before recombination. Thus, it makes sense to ask about DE large
scale rest frame given that it is supposed to interact very weakly
with the rest of particles of the universe. Moreover, in
\cite{flows} it is shown that the presence of a moving DE
component at the epoch when photons decouple from baryons could
straightforwardly account for the observed dark flow of matter
with respect to radiation. In this paper we show how the presence
of a moving DE component would affect the CMB temperature
fluctuations.

\section{Cosmology with moving fluids}

We consider a universe filled with four homogeneous perfect
fluids, namely: baryons (B), radiation (R), dark matter (DM) and
dark energy (DE), so that the total energy-momentum tensor reads
$T^{\mu\nu}=\sum_\alpha\left[(\rho_\alpha+p_\alpha)u_\alpha^\mu
u_\alpha^\nu-p_\alpha g^{\mu\nu}\right]$, where $\alpha=B,R, DM,
DE$ and $u_\alpha = \gamma_\alpha^2(1, v_\alpha)$ are the
velocities of the fluids with $\gamma_\alpha$ a normalization
factor determined by $u_{\alpha\mu}u_\alpha^\mu=1$. Besides, every
fluid satisfies a barotropic equation of state: $p_\alpha =
w_\alpha\rho_\alpha$. We shall also study the case in which DE is
a null fluid (a fluid moving at the speed of light) for which the
previous energy-momentum tensor is still valid, but with
$u_{DE\mu}u_{DE}^\mu = 0$. We can see that, in general, the
$(^0\;_i)$ component of Einstein equations yields the following
algebraic relation:
\begin{eqnarray}
g_{0i}\equiv
S_i=\frac{\sum_\alpha\gamma_\alpha^2(\rho_\alpha+p_\alpha)v_{i\alpha}}
{\sum_\alpha\gamma_\alpha^2(\rho_\alpha+p_\alpha)},\label{ccmc}
\end{eqnarray}
The combination $\rho_\alpha +p_\alpha$ appearing in this
expression is usually interpreted in General Relativity as the
density of inertial mass of the fluid so that we can interpret
$\vec{S}$ as the cosmic center of mass (CCM) velocity. Notice that
a pure cosmological constant with equation of state $w_{DE} = -1$
has vanishing inertial mass and, therefore, does not contribute to
the CCM velocity.

Since matter and radiation were coupled in the early universe,
their velocities must lie along the same direction. On the other
hand, in the CCM rest frame, where $\vec{S}=0$, the DE velocity
must lie along the opposite direction according to (\ref{ccmc}) so
that we have axial symmetry around the direction given by the
velocities and, if we choose the velocities lying along the
$z$-axis, the metric will be given by
$ds^2=dt^2-a_{\perp}^{2}(dx^2+dy^2)-a_{\parallel}^2dz^2$. However,
since observations show that the anisotropy of the universe, if
any, is small, we can assume that $a_{\perp ,\parallel} = a(1
+\delta_{\perp ,\parallel} )$, with $a$ the usual scale factor and
$\delta_{\perp ,\parallel} \ll 1$. Moreover, one can define the
{\it degree of anisotropy} by means of $h =
2(\delta_{\parallel}-\delta_{\perp})$, whose solution according to
Einstein equations to first order in $\delta$'s is given by
\cite{quadrupole}:
\begin{equation}
h=6\int_{a_*}^a\frac{1}{\tilde{a}^4}\left[\int_{a_*}^{\hat{a}}\hat{a}^2
\sum_\alpha(\rho_\alpha+p_\alpha)\sinh^2\theta_\alpha\frac{d\hat{a}}{\sqrt{\sum_\alpha\rho_\alpha}}\right]
\frac{d\tilde{a}}{\sqrt{\sum_\alpha\rho_\alpha}}.\label{hsol}
\end{equation}
where we assume that the universe is initially isotropic, i.e.
$h(a_*) = \dot{h}(a_*) = 0$ and $\theta_\alpha$ is the rapidity,
defined by $\tanh \theta_\alpha = a_\parallel v_\alpha\equiv
V_\alpha$. When the velocities are small we can approximate
$\sinh^2\theta_\alpha \simeq V_\alpha^2$ so that we conclude that
the first contribution to the anisotropy is of second order in the
velocities. The solution given by (\ref{hsol}) is still valid when
we consider a null fluid if we set $\sinh^2\theta_{DE} =1$ and, in
such a case, the dominant contribution to the degree of anisotropy
comes from the null fluid and we can neglect the rest of
components in the sum. Apart from Einstein equation, we also need
the energy and momentum conservation equations for each fluid
which can be solved in two limits:
\begin{itemize}
\item Slow-moving fluids. When the velocities are small, the
energy densities are unaffected by the motion and we have the
usual evolution: $\rho_\alpha = \rho_{\alpha 0}a^{-3(w_\alpha +
1)}$ whereas the velocities evolve according to $V_\alpha =
V_{\alpha 0}a^{3w_\alpha-1}$.

\item Fast-moving fluids. In the ultrarelativistic limit, the
physical velocity remains constant and the energy density evolves
according to $\rho_{\alpha} = \rho_{\alpha 0}a^{-2(1 + w)/(1 -
w)}$.
\end{itemize}
For a null fluid the energy-momentum conservation equations allow
us to obtain the solutions $p_N = p_{N0}$ and $\rho_N =
\rho_{0N}(a_\parallel a_\perp)^{-2}-p_{N0}$ without assuming any
particular equation of state. Moreover, if we require the energy
density to be positive at all times we have to impose $p_{N0} < 0$
so that a null fluid behaves as radiation during the early
universe and as a cosmological constant at late times.

\section{Contributions to the CMB}
The motion of the fluids has effects on CMB temperature
fluctuations through the Sachs-Wolfe effect. To first order in the
velocities it only affects the dipole according to the following
expression \cite{dipole}: $(\delta T/T)_{dipole} =
\vec{n}\cdot(\vec{S} - \vec{V})^{dec}_0$ where $0$ and $dec$
denote present and decoupling times respectively. Thus, the dipole
must be interpreted as a Doppler effect due to the relative motion
of the observer with respect to the cosmic center of mass, and not
just with respect to the CMB rest frame. That way, an observer who
measures a vanishing dipole would be at rest with respect to the
cosmic center of mass. This cosmological dipole contribution can
be identified with the anomalous dipole detected in
\cite{darkflow} that gives rise to the dark flow of matter. In
fact, its observed amplitude can be related to the present
velocity of DE which, assuming $w_{DE}^0\simeq -0.97$, can be
estimated to be $v_{DE}(t_0)\sim 1$km$/$s \cite{flows}.

In the case of small anisotropy, it is possible to see that the
temperature fluctuation has two contributions \cite{Bunn}: $\delta
T_T = \delta T_I + \delta T_A$, where $\delta T_I$ is the
fluctuation produced during inflation and $\delta T_A$ is due to
the anisotropy generated by the motion of the fluids, which is
given by \cite{quadrupole}: $(\delta T_A/T) = 2|h_0 -
h_{dec}|/(5\sqrt{3})$. For an arbitrary direction of the
velocities $(\theta, \phi)$ and assuming a statistically isotropic
distribution for inflation fluctuations, we get $(\delta T)_T^2 =
(\delta T)_A^2 + (\delta T)_I^2 + f(\theta, \phi, \alpha_i)\delta
T_A\delta T_I$ where $\alpha_i$ are random phase factors coming
from inflation and $f$ is a function satisfying $|f | \leq 0.98$
\cite{Campanelli}. Then, the total quadrupole lies between
$(\delta T)_-^2$ and $(\delta T)_+^2$ with $(\delta T)_\pm^2 =
(\delta T)_A^2 + (\delta T)_I^2 \pm 0.98 \delta T_A\delta T_I$. On
the other hand, the observed quadrupole is in the range $91.51
\mu\K^2 \leq (\delta T)^2_{obs} \leq 406.48 \mu\K^2$, (including
the cosmic variance). Thus, if we assume that inflation
contribution agrees with the central measured value, i.e.,
$(\delta T)_I^2 = 247 \mu\K^2$, the anisotropic contribution
should satisfy $(\delta T)_A^2 \lsim 1254 \mu\K^2$, which defines
the allowed region for the dark energy models. However, standard
inflation predicts a larger value: $(\delta T)_I^2 = 1252
\mu\K^2$. In such a case, if $247.90 \mu\K^2 \lsim (\delta T)_A^2
\lsim 2883.80 \mu\K^2$  then the fluids motion could explain the
low observed quadrupole for certain values of dark energy velocity
and phase factors.
\subsection{Model examples}

\begin{itemize}

\item {\it Constant equation of state}. For a model with constant
equation of state close to $-1$ (but different from $-1$) we find
that the velocities are extremely small so that all the fluids are
nearly at rest and the effect on the quadrupole is completely
negligible.

\item {\it Scaling models}. In these models the equation of state
of DE mimics that of the dominant component of the universe
eventually exiting from this regime and joining into one with
constant equation of state close to $-1$. The quadrupole generated
by scaling models is fixed by two parameters: the initial DE
fraction $\epsilon$ and its initial velocity $V^*_{DE}$, and for
small velocities it is given by $\delta T_A/T\simeq 0.44 \epsilon
(V^*_{DE})^2$. Then, if we take into account the bounds obtained
above we get that the allowed region is: $\epsilon (V^*_{DE})^2
\lsim 2.9\times 10^{-5}$. Moreover if $1.3\times 10^{-5} \lsim
\epsilon (V^*_{DE})^2 \lsim 4.3\times 10^{-5}$ these models could
explain the low quadrupole.

\item {\it Tracking models}. In tracking models the equation of
state of DE is initially close to $1$ so that the velocity grows
as $a^2$ until it reaches the speed of light and then it remains
constant according to the solution for fast moving fluids. In that
regime, the energy density starts falling extremely fast and
eventually becomes completely negligible, giving rise to a fluid
of vanishing energy density moving at the speed of light.
Therefore, we can conclude that tracking models are unstable
against velocity perturbations.

\item {\it Null Dark Energy}. When DE is described by a null
fluid, the quadrupole is fixed just by $\epsilon$. For small
velocities, the quadrupole is approximately given by $\delta T_
A/T \simeq 2.58\epsilon$ and the constraints on the anisotropic
contribution lead to following allowed region: $\epsilon \lsim
5\times 10^{-6}$. Again, if $2.2\times 10^{-6} \lsim \epsilon
\lsim 7.6\times 10^{-6}$, these models could explain the observed
quadrupole with the standard contribution from inflation.
\end{itemize}

\section{Conclusions}
In this work we have shown how a moving DE fluid can generate
large-scale anisotropy starting from an isotropic universe and how
this can affect the CMB temperature fluctuations. In particular,
we have seen that in such a case, the CMB dipole is due to the
relative motion of the observer with respect to the cosmic center
of mass and, besides, it could explain the observed dark flow of
matter. Concerning the quadrupole, we obtain bounds on the
anisotropic contribution by comparing with observations and found
that scaling and null DE models succeed in explaining the low
measured value starting from the standard contribution from
inflation.


\begin{theacknowledgments}
We would like to thank the organizing committee of the Spanish
Relativity Meeting ERE 2008 for their hospitality and excellent
organization. This work has been supported by DGICYT (Spain)
project numbers FPA 2004-02602 and FPA 2005-02327, UCM-Santander
PR34/07-15875, CAM/UCM 910309 and MEC grant BES-2006-12059.
\end{theacknowledgments}






\end{document}